\documentclass[conference]{IEEEtran}
\IEEEoverridecommandlockouts

\usepackage[T1]{fontenc}
\usepackage{cite}
\usepackage{amsmath,amssymb,amsfonts}
\usepackage{algorithmic}
\usepackage{graphicx}
\usepackage{textcomp}
\usepackage{xcolor}
\usepackage{booktabs}
\usepackage{multirow}
\usepackage{todonotes}
\usepackage{url}
\usepackage{hyperref}
    
\begin{document}

\title{

Human and Machine: How Software Engineers Perceive and Engage with AI-Assisted Code Reviews Compared to Their Peers

}

\author{\IEEEauthorblockN{1\textsuperscript{st} Adam Alami}
\IEEEauthorblockA{\textit{M{\ae}rsk Mc-Kinney M{\o}ller Institute} \\
\textit{University of Southern Denmark}\\
}

\and

\IEEEauthorblockN{2\textsuperscript{nd} Neil Ernst}
\IEEEauthorblockA{\textit{Department of Computer Science} \\
\textit{University of Victoria}\\
}
}

\maketitle

\begin{abstract}

The integration of artificial intelligence (AI) continues to increase and evolve, including in software engineering (SE). This integration involves processes traditionally entrusted to humans, such as coding. However, the impact on socio-technical processes like code review remains underexplored. In this interview-based study (20 interviewees), we investigate how software engineers perceive and engage with Large Language Model (LLM)-assisted code reviews compared to human peer-led reviews. In this inherently human-centric process, we aim to understand how software engineers navigate the introduction of AI into collaborative workflows. We found that engagement in code review is multi-dimensional, spanning cognitive, emotional, and behavioral dimensions. The introduction of LLM-assisted review impacts some of these attributes. For example, there is less need for emotional regulation and coping mechanisms when dealing with an LLM compared to peers. However, the cognitive load sometimes is higher in dealing with LLM-generated feedback due to its excessive details. Software engineers use a similar sense-making process to evaluate and adopt feedback suggestions from their peers and the LLM. However, the LLM feedback adoption is constrained by trust and lack of context in the review. Our findings contribute to a deeper understanding of how AI tools are impacting SE socio-technical processes and provide insights into the future of AI-human collaboration in SE practices.

\end{abstract}


\section{Introduction}\label{sec:introduction}

\noindent The integration of artificial intelligence (AI) is growing in many processes, including software engineering (SE) \cite{fan2023large}. However, for decades and despite various practices being automated, SE remains a human activity. The integration of AI into a socio-technical process such as SE, historically dominated by human control, interactions, and decision-making, may imply a paradigm shift in the way software is engineered.

Research and products aiming to leverage AI for software engineering are targeting different core SE activities. Research efforts have shown interest in requirements engineering, e.g., \cite{mich2002nl,ninaus2014intellireq,del2015multi}, software design, e.g., \cite{jiao2004automated,rodriguez2016artificial}, construction, e.g., \cite{ling2016,yu2018spider} and testing, e.g., \cite{zaeem2014automated,schneid}.

Machine learning (ML) technologies have opened up the possibility to automate difficult SE tasks, traditionally entrusted to humans. Advances in neural network architectures, such as recurrent neural networks and transformers \cite{vaswani2017attention}, have been used to advance the state-of-the-art for many difficult automated software engineering tasks \cite{dehaerne,fan2023large}. These technologies, among others, have contributed to an array of commercial products such as Tabnine \cite{TabnineA6}, CodeX \cite{codex}, Github's Copilot \cite{GitHubCo57} and ChatGPT.

While these products are task specific, i.e., AI pair programming, Large Language Models (LLMs) like GPT-4 have emerged as powerful multi-task products, capable of not only generating code but also assisting in other software development activities, including documentation, debugging, design suggestions, and code review \cite{fan2023large}. The versatility and broad applicability of LLMs have the potential to fundamentally alter the landscape of software engineering by either reducing the reliance on human intervention or support in tasks that were previously considered too complex for automation \cite{svyatkovskiy2020,zohair2018future}.

Although the products landscape designated to LLM-supported SE and the enthusiasm for their adoption are increasing \cite{fan2023large,Octovers}, our understanding of the impact of this shift on the social and human dynamics within SE environments remains limited \cite{anthony2023collaborating}. SE is inherently a socio-technical practice \cite{baxter2011socio}. The engineering process is entwined with collaboration, human judgment, emotional interactions, and decision-making \cite{kalliamvakou2017makes}. As we witness a potential paradigm shift—moving from human-dominated processes to AI-supported workflows—it is important to investigate how these changes affect the human and social elements of SE. 

Most of human SE activities are collaborative. Thus, the integration of AI into the already complex human and social workflows necessitates a thorough understanding of the implications. Specifically, the introduction of AI in practices traditionally governed by human expertise raises questions about the human-AI engagement and behavioral responses of practitioners like software engineers. For example, Malone et al. suggest that the most challenging changes in AI integration are not computers replacing humans but rather people and computers working together as integrated ``superminds'' \cite{malone2020artificial}. To effectively harness the potential of these ``superminds'' \cite{malone2020artificial}, it is crucial to understand how software engineers engage with and respond to AI tools in their workflows. Therefore, we propose:

\medskip

\noindent \textbf{RQ:} How do software engineers perceive and engage with LLM-assisted code reviews compared to their peers?

\medskip

Engagement, in the context of our study, is the ways and the extent to which software engineers actively interact with, respond to, and incorporate feedback from the review process. We opted for code review as a SE process, because it is inherently a human-centric process that relies on collaboration, judgment, and communication. This makes it ideal to evaluate how engineers experience the engagement in an LLM-supported alternative to a process traditionally characterized by human-intensive interactions.

To investigate our \textbf{RQ}, we set an interview study (20 interviewees) anchored in tangible review experiences. Prior to the interviews, we asked our participants to submit a sample of their own authored code. Then, we assigned two to three human reviewers to every author and asked them to submit written reviews. We used the reviews during the interviews to prompt our interviewees to reflect on their experiences, perceptions, and emotional reactions using recent and concrete experiences. In the second part of the interview, we shared an LLM-generated review of their authored code to explore how their reactions and engagement differ when interacting with AI-generated feedback as opposed to human-generated feedback.

By juxtaposing these insights, we aimed to understand and capture any deviations in engagement in the case of LLM-assisted reviews compared to peers. This allowed us to identify challenges and constraints in the integration of AI tools into collaborative SE processes. This understanding may inform the development of more effective and harmonious AI-human collaborative workflows in the future.

We contribute to AI integration in SE literature by identifying the dimensions of engagement in code review, which include cognitive, emotional, and behavioral responses to feedback, and how the introduction of LLMs could influence hese engagement attributes.

\section{Related Work}\label{sec:related}

While code review is part of our research context, it is not the primary focus of our investigation. We used code review as a lens through which we evaluated a broader and critical issue of how software engineers perceive and engage with AI tools in a socio-technical process. In particular, we examine how engineers respond to AI-generated feedback compared to feedback from peers. Therefore, the scope of our related work is to draw on studies focusing on human-AI collaboration and how software engineers engage with AI in SE processes. We set our results in the context of existing literature, including code reviews, in a discussion section (\S \ref{sec:discussion}).

Human-tool integration before the advent of AI (i.e., LLM-based assistants) often took the form of reporting automated build and test results or other static analysis checks (e.g., linting or security scans). A simple example is the output of a compilation run and the resulting error messages. Even this relatively simple set of tool results can be difficult for humans to understand and work with~\cite{Denny2021}. Static analysis results might take the form of reporting that a particular code file contains known security issues (e.g., possible SQL injections). This is a considerably more actionable and explicable report than the detailed AI-generated code reviews we consider in this paper. But even these simpler reports are challenging to integrate into human decision-making. For example, Johnson et al. showed that reporting the factual outputs of the analysis is not sufficient for developers to adopt a static analysis tool~\cite{johnson2013}.

A common medium for human-tool integration is a bot, typically serving as a textual/chat interface to the underlying static analyzer. For example, Repairnator was a bot for GitHub that was an interface to a sophisticated program repair tool~\cite{Monperrus2019}. The bot's designers found that without \emph{explanation} it was hard or impossible to get humans to accept the changes~\cite{Monperrus2019b}.

In a wider study, researchers examined bot trust and autonomy and found that individuals have varying levels of trust in the bot's recommendations~\cite{ghorbani2023autonomy}. That paper recommended that bot interaction characteristics should be customizable. In addition, \emph{how the bot appears} (i.e., avatar, language, name) can have a big impact on human acceptance: merely revealing that something is a bot can dramatically change human perceptions~\cite{Murgia2016}.

These perceptions may be changing as LLM capabilities (and surrounding hype) make AI assistance more commonplace. Nonetheless, Al Haque et al. have identified several challenges, including that the AI-generated content was found to be overly polite, overly detailed, and lacked trustworthiness~\cite{alhaque}. To date, most of the work on LLMs and coding, including code review, has focused on the technical aspects of LLMs rather than human interaction issues. Technical work looks at creating datasets of pull requests and code problems~\cite{Li2022}, examining how to improve fine-tuning, e.g., LLaMA-Reviewer~\cite{Lu2023}, and exploring different architectures for the underlying neural networks~\cite{Tufano2022}.

Lack of specific context is a challenge for AI tools. Developers get annoyed when the AI does not understand the locality of their codebase, although for boilerplate or generic code, developers appreciate saving keystrokes~\cite{Liang2024}. As Panichella et al. point out~\cite{Panichella2020}, since code review tasks themselves are frequently changing, e.g., as the organization develops a new understanding of concurrency, statically trained LLMs may miss evolving changes. 

In sum, these studies show that achieving effective human-tool integration is about much more than functionality \cite{Li2022,Lu2023,alhaque}. Tool design must also address human aspects such as trust, interpretability, and engineers’ preferences for communication style and content \cite{ghorbani2023autonomy,Murgia2016}. Our study contributes to this body of work by identifying potential shifts in software engineers engagement in response to AI-driven code review.

\section{Methods}\label{sec:methods}

\noindent We opted for an interview study to investigate our \textbf{RQ} objectives. This choice is better suited when the phenomenon under study requires uncovering complex social and behavioral nuances that influence human judgment and interactions in socio-technical processes \cite{kvale2009interviews,baxter2011socio}. Interviews are also a good tool to access perspectives and insights ingrained deeply in personal beliefs, attitudes, and behaviors \cite{larkin2021interpretative,seidman2006interviewing}. By tapping into these core personal and deep-seated cognitive constructs \cite{larkin2021interpretative,seidman2006interviewing}, we aim to collect data in line with our study objectives.

We designed a research process to capture real-life practices, yet within the constraints inherent to a research environment. For example, our participants remained anonymous to each other, opposed to a professional context where authors and reviewers are known to each other. Upon the completion of our recruitment and selection process, we asked our participants to submit a code they had authored. Then, we assigned to each author two to three reviewers (drawn from other participants in the study) and asked them to submit a written review as if they were to conduct a peer review in their professional contexts. Prior to the interviews, we shared the peer reviews with the authors. We asked the participants permission to prompt ChatGPT to review their code and inform them that we will share the LLM-generated reviews in the interview for discussion.

In the first part of the interview, we used the reviews as catalysts to prompt the engineers to explain and elaborate further on their responses to our questions. In the second part, we shared an LLM-generated review of the same code authored by the interviewees to prompt them to share their attitudes and understand how they would engage with it compared to human-authored reviews. This approach ensures that we capture the nuanced differences in how software engineers engage with and respond to both human and LLM-generated feedback. First, we grounded the discussion in their real-world practices and the study's review experience. Then, we prompted them to contrast their earlier claims when reviews are AI-generated. This dual-method strategy, combining data in both engagement with peer-conducted and AI-generated reviews, aligns with the study objectives. The process was explained to participants during the recruitment.

\begin{table*}[th!]

  \begin{center}
    \footnotesize
    \caption{Interviewees' charachterestics.}
    
    \vspace{-0.5cm}
    
    \label{tbl:population}
    \renewcommand\arraystretch{0.80}
        
    \begin{tabular}{l|p{2.5cm}|c|c|p{2.8cm}|p{1.8cm}|p{2cm}|p{1.5cm}}
      \hline
      \textbf{\#} & \textbf{Role} & \textbf{Exp.} & \textbf{Gender} & \textbf{Industry sector} & \textbf{Language} & \textbf{Reviewers} & \textbf{Country}\\
      \hline
    
        P1 & Sr. Software Engineer & 6-10 years & Male & IT Services & JavaScript & P10, P19, \& P20 & Germany\\
        P2 & Sr. Software Engineer & 6-10 years & Male & Consulting & Java & P1, P3, \& P19 & UK\\
        P3 & Sr. Software Engineer & >10 years & Male & Telecommunication & Python & P4, P12, \& P18 & UK\\
        P4 & Software Engineer & 3-5 years & Male & Telecommunication & Python & P3, P12, \& P20 & Germany\\
        P5 & Software Engineer & >10 years & Male & IT Services & C\# & P11 \& P17 & USA\\
        P6 & Software Engineer & 6-10 years & Non-binary/third gender & Finance & Java & P6, P11, \& P14 & Netherlands\\
        P7 & Tech Lead & >10 years & Male/ & Finance & Java & P11 \& P14 & Canada\\
        P8 & Software Engineer & 3-5 years & Male & Aviation & Python & P9, P17, \& P20 & Spain\\
        P9 & DevOps Engineer & >10 years & Female & Health Care & Python & P8, P13, \& P20 & UK\\
        P10 & Software Engineer & 1-2 years & Male & IT Services & Python & P3, P12, \& P18 & UK\\
        P11 & Software Engineer & 3-5 years & Female & Finance & Java & P6 \& P7 & South Africa\\
        P12 & Sr. Software Engineer & >10 years & Female & Finance & Python & P4, P12, \& P19 & UK\\
        P13 & Sr. Software Engineer & 6-10 years & Male & IT Services & TypeScript & P16 \& P17 & Ireland\\
        P14 & Software Engineer & 6-10 years & Male & IT Services & Java & P5, P8, \& P13 & Portugal\\
        P15 & Sr. Software Engineer & 3-5 years & Female & Government Services & Python & P16 \& P18 & Italy\\
        P16 & Software Engineer & 1-2 years & Male & IT Services & JavaScript & P6, 13, \& P21 & Portugal\\
        P17 & Software Engineer & 3-5 years & Male & Media \& Entertainment & Python & P5, P8, \& P9 & Uk\\
        P18 & Software Engineer & 6-10 years & Male & IT Services & C++ & P2, P12, \& P15 & UK\\
        P19 & Sr. Software Engineer & 6-10 years & Female & IT Services & Java & P2 \& P10 & Germany\\
        P20 & Software Engineer & 3-5 years & Non-binary/third gender & IT Services & JavaScript & P1 \& P10 & Ireland\\
    
     \bottomrule
     
    \end{tabular}
   
  \end{center}
  
  \vspace{-0.8cm}
  
\end{table*}

\subsection{Interviewee recruitment \& selection}

\noindent To recruit our interviewees, we used Prolific\footnote{\url{https://www.prolific.co/}}, a research market platform. Given that the platform does not verify nor evaluate self-reported skills \cite{alami2024you}, we carried out a pre-screening process to qualify our potential participant, following the guidelines suggested by Alami and colleagues \cite{alami2024you}. 

Alami et al. recommend using an iterative and controlled prescreening process, using task-oriented questions that go beyond theoretical understanding to help filter genuine engineers from those who may rely on external resources, such as Googling or prompting an LLM for answers \cite{alami2024you}. 

In the pre-screening survey, we used a programming task, a critical-thinking question, and a prompt for participants to share a problem-solving scenario from their own experience. We iteratively and manually assessed the survey answers \cite{alami2024you}. First, we evaluated the answers for AI-generated content using ChatGPT 4.0. Subsequently, we manually evaluate the content's quality and coherence \cite{alami2024you}. We capped the number of participants to 500, and after the prescreening process, we ended up with 353 qualified engineers. Then, we ran a final pre-selection survey where we asked for further demographic data, programming language skills, and whether the participants were willing to participate in an interview study. We limited this pre-selection to 100, and 76 agreed to participate in the interviews. This 2-phased approach had different objectives. While in the initial prescreening, we focused on evaluating potential participants' skills, i.e., are they genuinely software engineers \cite{alami2024you}, in the second pre-selection, we wanted to know whether those qualified software engineers are willing to take part in an interview study and the programming languages they may submit their code and review other participants' codes. We paid \textsterling0,50 for the initial prescreening and pre-selection surveys and \textsterling60,00 for the participation in the interview. The selection process took place in the first and second weeks of August 2024.

Table \ref{tbl:population} documents our participants' characteristics, providing their current roles, experience levels, gender, industry sectors, and the programming languages they opted to submit their code in. The column ``Reviewers'' lists the reviewers assigned to each participant. We managed to assign to each participant a minimum of two reviews. Even though we aimed for three reviews per participant to align with industry and open-source community practices \cite{alami2020foss,alami2021pull}, we did not manage to meet this expectation due to the varying programming skills and language preferences among participants, which limited the pool of suitable reviewers for certain code submissions. In assigning reviewers to authors, we aimed for a diverse range of experiences and backgrounds. For example, P1, a senior software engineer with an experience of 6–10 years, was assigned reviewers with experiences ranging from less than two years to ten years. This diversity in reviewers' experience levels was intentionally designed to enrich our data with varying levels of expertise and perspectives and feedback from both novice and seasoned reviewers. We also aimed to be diverse in gender and country in the reviewers selection. For example, two females and one male from two different countries reviewed P18.

\subsection{Data collection}

\begin{table}[th!]
  \footnotesize
  \caption{Key parts of the interview guide}%

  \vspace{-0.2cm}
  
  \renewcommand\arraystretch{1.3}

  \begin{tabular}{p{8.6cm}}
     \hline
     \textbf{Introduction}\\
     \midrule
    
      Can you please introduce yourself? (Include educational background, current role, and years of experience in software development.)\\
      
      \hline
      \textbf{Section I: Approach to Reviewing Code }
      \\\midrule
 
        Are there any coding standards or best practices that you kept in mind during the reviews we assigned to you?\\
        What did you consider most when you wrote your feedback for the assigned code?\\
    
     \hline
     \textbf{Section II: Engagement with peers’ feedback on authored code }
     \\\midrule

        How do you typically feel when you receive feedback on your code from peers?\\
        In the feedback you received from the other participants, what elements of the feedback do you find most useful or valuable? And why?\\

     \hline
     \textbf{Section III: Engagement with LLM-generated review }
     \\\midrule

        How did you feel when you first read the LLM-generated review of your code?\\
        Would you respond and react to the LLM-generated feedback the same way as you do to human's feedback?\\

    \hline
     \textbf{Conclusion}
     \\\midrule

        Based on your experiences in this study, how would you summarize the main differences between peer and LLM-generated code reviews?\\
        Is there anything else you would like to share about your experience in this study or with code reviews in general?\\

    \bottomrule
    
  \end{tabular}
  
  \label{tbl:guide}
  
\vspace{-0.65cm}

\end{table}

\begin{table*}[ht!]
\footnotesize
    \caption{Example of Pattern Codes and their corresponding First Cycle codes}
    \vspace{-0.2cm}

    \label{tbl:themes}
    \renewcommand\arraystretch{1.8}
    
    \begin{tabular}{p{4.1cm}p{2.8cm}p{10cm}}
    \toprule
      \textbf{Pattern codes} & \textbf{First Cycle codes} & \textbf{Examples from the data}\\
      \hline
      
      \multirow{8}{*}{\textbf{Reviewer context}} & Peers' seniority  & \emph{``... if it's a junior developer, I just have more suggestions about things that I spot in the code. Where is if it's a senior developer, sometimes I won't have as many suggestions, and I'll just let them know maybe one or two things. But other than that, I'll just let them know looks good to me, and I'll give it a thumbs up for approval''} (P5).\\
      
      & Familiarity with peers & \emph{``I know the team very well, the team knows me very well ... when you're in that kind of environment, you kind of, you know, you get straight to the point, because, you know, you don't want to, you don't want to waste their time ... I don't really put the positive feedback, because it was like, you know, they probably, you know that it's good''} (P2). \\
      
      & Trust in LLMs & \emph{``... with chatgpt, I can't be like I would 100\% validate ... I would not go like, just blindly trust it. And with my peers, I would at least have a general idea of, like, how much knowledge they have and how, how well they are delivering feedback''} (P4). \\
      
      \hline
      
      \multirow{5}{4.1cm}{\textbf{Engagement}} & Cognitive  & \emph{``... [The review received from] number six [i.e., P6] was the most, easiest for me to digest, and I felt like the tone wasn't really too harsh, or I think it was very professional''} (P7). \\
      
      & Emotional  & \emph{``It feels nice [to receive helpful feedback]. It feels like he genuinely cares for the code I produce, and he wants to help me do well''} (P10).  \\ 

      & Behavioral & \emph{``I think it [peer reviews] help to improve myself. So always even be a better version of myself''} (P15). \\
      
     \bottomrule
     
    \end{tabular}
    
  \vspace{-0.5cm}
  
\end{table*}

\noindent Our research design sought to ground the data collected during the interview in concrete experiences by our participants. By anchoring the interview discussions in reviews received from other participants, we grounded the data in personal, recent, and close experiences. To this end, prior to the interviews, we shared the reviews with all participants and asked them to read and reflect on them in preparation for the interview.

The interview consisted of two parts. In the first part, we focused on the engagement with human peer's reviews, and in the second part, we shared a ChatGPT 4.0-generated review with the participants and asked them to share their perceptions and reactions to the LLM-generated feedback on their code. We opted for ChatGPT for its wide accessibility and familiarity among a diverse audience. To generate ChatGPT reviews, we used this prompt: ``You are an expert of [the programming language]. Provide a thorough review of the attached code.'' This prompt design aimed to generate expert-level feedback, relatable to what a human reviewer might offer \cite{brown2020language}. Such feedback is what developers would expect from knowledgeable peers \cite{brown2020language}. The intent of this prompt design is also to create a realistic basis for comparing human and AI-generated reviews. For example, a prompt that lacks specificity or context may generate unrealistic and unrelatable feedback \cite{liu2023pre}.

Our interviews were semi-structured. Although we designed an interview guide, its purpose was to guide the conversation while allowing fluidity by prompting the interviewee to elaborate and explain further based on their responses. We also used examples from the reviews they received to engage them in deeper reflection of how they perceived the comments they received either from the LLM or other participants. Table \ref{tbl:guide} illustrates examples of the questions we asked. The full guide is available in the shared documents package (Sect. \ref{sec:replication}).

All interviews were conducted using Zoom and lasted 40-60 minutes, with a total of 17h12min of audio. The audio generated a total of 271 pages verbatim after transcribing, an average of approximately 14 pages per interview. The first author conducted the interviews in the first and second weeks of September 2024. We used Otter.ai\footnote{\url{https://otter.ai/}} an online transcription tool to transcribe the audio recordings.

\subsection{Data analysis}

\noindent We adopted a constructivist stance \cite{jw1998qualitative}, which aligns with our objective to understand software engineers' perceptions and  experiences with AI-generated feedback compared to their peers. By adopting this stance, we recognize that knowledge and meaning are actively constructed by individuals through their experiences and interactions \cite{jw1998qualitative}.

We followed Miles et al. \cite{miles2014qualitative} recommendations for the analysis process. First, we conducted a preliminary \emph{First Cycle} analysis. In this early stage of the analysis, we selected ``chunks'' of data that are pertinent to our RQ and assigned them codes \cite{miles2014qualitative}. Using an inductive strategy, this condensing yet analytical exercise allowed us to reduce the data corps into meaningful concepts \cite{miles1984qualitative}.

In the \emph{Second Cycle} that followed, we synthesized all the \emph{First Cycle} codes into ``Pattern codes'' \cite{miles2014qualitative}. In this integrative activity, we looked for patterns across the previous phase codes by identifying recurring themes that linked different codes together; either have similarities, form a process, or their relations form a cohesive construct by complementing each other \cite{miles2014qualitative}. This process allowed us to develop higher-level constructs and refine our understanding of the underlying phenomena.

Table \ref{tbl:themes} documents an example of some of our Pattern Codes and their corresponding \emph{First Cycle} codes. Each pattern code was counted once per interview, regardless of the number of times it was mentioned within a particular interview. The final column provides examples from the data.

The first and second coding cycles were conducted by the first author. Then, iteratively, the second author reviewed the codes and Pattern Codes, provided comments, and proposed new and alternative codes, until a consensus was reached. The first author then revised and offered a final set of codes and Pattern Codes. This ``reliability check'' \cite{miles2014qualitative,creswell2016qualitative}, allowed us to validate our coding judgments and settle our discrepancies, resulting in more trustworthy interpretations.

We monitored data \emph{saturation} \cite{morse2004theoretical,aldiabat2018data} throughout our analytical process. We commenced the analysis as soon as the interview transcripts became available. Then, iteratively, as data and the analytical outputs became available, we compared data and emerging themes to assess saturation points \cite{morse2004theoretical}. By switching back and forth between data collection and analysis \cite{bowen2008naturalistic}, we managed to observe when our Pattern Codes reoccur strongly in the data, hence reaching saturation. We documented this process in our shared documents package.

Upon the completion of the analysis, we organized a \emph{member checking} \cite{birt2016member} activity to collect feedback from our interviewees on our findings (see Sect. \ref{sec:validity} for further details).

\subsection{Ethics and informed consent}

\noindent All interviewees were asked to read a formal consent request on the first page of the pre-selection survey and asked to agree or disagree to participate. The consent covered anonymized data sharing in public repositories and manuscripts, the voluntary nature of participation, withdrawal from the study, the procedures for ensuring confidentiality, and other ethical considerations as per the authors' university policies. Ethical approvals, as per the authors' university requirements, were obtained prior to the study commencing.

\subsection{Shared documents package}\label{sec:replication}

\noindent In this package\footnote{ Interview transcripts will be published on acceptance.}, we share code snippets and reviews submitted by all participants, the interview pre-screening questionnaire, ChatGPT-generated reviews, member checking questionnaire and responses, and the interview guide. The package can be found \href{https://doi.org/10.5281/zenodo.14000259}{here.}\footnote{\url{https://doi.org/10.5281/zenodo.14000259}.}

\section{Findings}\label{sec:findings}

\begin{figure*}[t]

\includegraphics*[trim=0.5cm 7.1cm 1.5cm 1cm, clip, width=.90\textwidth]{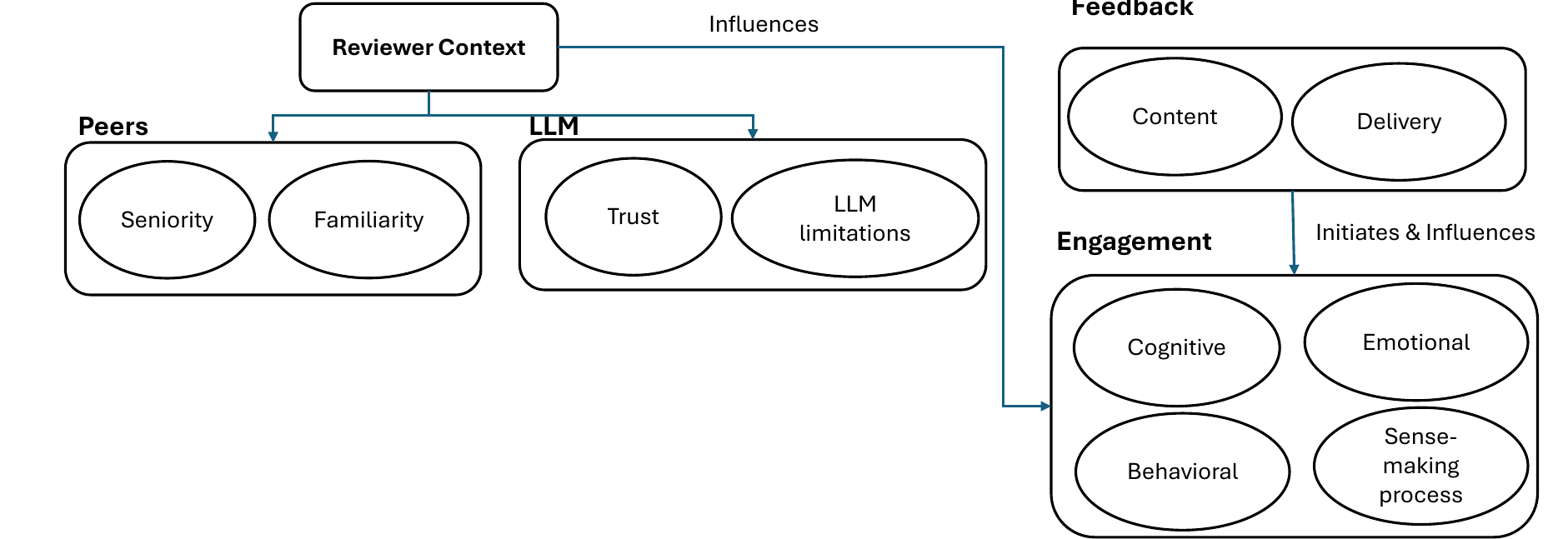}

\caption{An abstract presentation of the findings, capturing engagement and how engineers respond to feedback. The line from \textbf{Reviewer Context} to \textbf{Peer} and \textbf{LLM} indicates two type of reviewers, and the ovals inside the rectangles are elements of the reviewer contexts.}

\vspace{-0.5cm}

\label{fig:fig1}

\end{figure*}

\noindent Recall, our \textbf{RQ} seeks to understand engagement in code review and how it may be impacted if we introduce LLM-assisted reviews. Engagement, in the context of our study, is the ways and the extent to which software engineers actively interact with, respond to, and incorporate feedback from the review process. \textbf{We found that engagement is multi-dimensional, spanning \emph{cognitive}, \emph{emotional}, and \emph{behavioral} responses}. While experiencing these responses, software engineers undertake a \emph{sense-making process} of the feedback in order to make a decision regarding its adoption. We also found that \emph{human-AI collaboration} in the context of code review is characterized by divergence in preferences regarding how feedback should be delivered. 

Engagement is initiated by receiving feedback from peers or the LLM (in the case of our study), which has two key components, \emph{content} and \emph{delivery}. Both influence how engineers become involved emotionally and cognitively with the feedback. For example, ``harsh'' (e.g., P17) and ``toxic'' (e.g., P2, P6, and P10) feedback evoke negative emotions, which in response prompt emotional regulation from the engineers as a coping mechanism. However, this emotional regulation becomes less relevant in the case of LLM-assisted review, as these tools use a positive and professional tone.

Engagement is also influenced by the \emph{reviewer context}. For peers, it is their seniority and familiarity with each other. For LLMs, it is the engineer's trust in LLMs capabilities and other inherent limitations, such as a lack of context relevant to the codebase. Figure \ref{fig:fig1} captures this dynamic. Below we elaborate on the themes of our findings.

\subsection{Feedback}

\noindent Feedback is the crux of a code review. It shapes the engagement process. Both the content and the delivery style determine the cognitive effort required by engineers; however, delivery style triggers more emotional and behavioral responses, either negative or positive.

For authors, the cognitive load appears to be low when the feedback is constructive or precise. For example, when P16 was asked to explain the rationale behind his direct and precise comments, he explained: \emph{``... you have to be direct and precise, otherwise people will not understand. But the more direct and logical, I guess it's easier for the programmer to understand what to do''} (P16).

When P8 was asked to explain his tone style in commenting on P9's code, he said: \emph{``... when I get code reviews that are like, super picky ... I think, like, the intention that you get from the code review makes you feel one way or another and makes you perform better or worse ... I won't be in the mood like, let's try to do it ... But if I get, like, a review that ... more constructive ... I will go try to do it my best ... open to more improvement''} (P8). This account shows resignation as an emotional response when the feedback is \emph{``super picky''} and comfort when \emph{``constructive.''}

The feeling of resignation or being out of the mood is demotivating for P8, \emph{``I won't be in the mood like, let's try to do it''} (P8). On the other hand, the feeling of comfort seems to have a contrasting effect: \emph{``I will try to do it my best''} (P8).

The introduction of LLM-assisted review showed that the cognitive load required to process LLM reviews seems higher; however, the emotional response appears to be less taxing compared to peers'. When P14 was asked to compare the comments from other participants and the LLM, he commented: \emph{``ChatGPT review will take more time and effort to analyze and review, and others are more straightforward; they identify the problem and suggest a solution for it. This [ChatGPT-generated review] will take more time and mental effort to analyze''} (P14). However, this could be improved with further fine-tuning of the prompt and training of future models based on a foundational model like ChatGPT.

Emotionally, engagement with LLM-generated reviews seems less taxing for our interviewees. When P8 was prompted to explain why he prefers ChatGPT reviews over some of the other participants, he explained: \emph{``... ChatGPT is also polite. So as long as it's polite, I don't really mind ... but a person may not be polite. So, yeah, I wouldn't mind [replacing peers with an LLM]. ChatGPT would never be picky''} (P8). This preference for ChatGPT seems to be rooted in its lack of emotional variability as opposed to humans, who can be ``not polite'' and ``picky.'' ChatGPT maintains a consistent positive tone, reducing the likelihood of P8 experiencing a taxing emotional experience during the review process. This was echoed in P10 answer to the same question, i.e., why preferring LLM, \emph{``I don't think it [ChatGPT] ever will be harsh ... it just made it easier to kind of accept what it was telling me''} (P10).

Feedback, whether constructive or overly critical, elicits varying emotional and behavioral responses, from resignation to motivation. Our data show that clear and constructive feedback tends to reduce cognitive load and fosters a sense of improvement for the code. The introduction of LLM-assisted reviews seems to offer a less emotionally taxing experience with their consistent positive tone, though potentially at the cost of increased cognitive effort.

\subsection{Engagement}

\noindent While experiencing their \emph{cognitive}, \emph{emotional}, and \emph{behavioral} responses to the feedback, software engineers engage in a \emph{sense-making process} to make a decision regarding its adoption. 

\paragraph*{Cognitive} The cognitive part of engagement refers to the mental resources and effort required to process feedback or review code. As reviewers, some engineers in our sample claim that reviewing large code changes is mentally demanding. P2 explains: \emph{``I like code review; when the tickets are small ... if they're doing big changes, your brain gets a bit tired ... sloppy, and you won't be able to catch everything''} (P2).

Our participants assumed the roles of code authors and reviewers for other participants' code and we designed the interview guide to prompt them about both roles. As authors, our participants claim that the cognitive load is lower when the feedback from peers is constructive or precise. P7 described the review he received from P6 \emph{``easy to digest''}, because he \emph{``felt''} that the tone was not \emph{``too harsh''} and \emph{``professional.''} P7’s experience demonstrates more efficient cognitive engagement; P6's feedback mitigated high cognitive load. The clarity and professionalism of P6's feedback made it easier for P7 to process, reducing his mental effort required to engage with it. This may suggest that feedback delivered in a constructive and respectful manner fosters both positive emotional responses and facilitates more efficient cognitive engagement.

Compared to LLM-assisted review, the cognitive load required to process the review is higher than that of peers' reviews. P14 states: \emph{``ChatGPT review will take more time and effort to analyze and to review, and others [participant's reviews] are more straightforward''} (P14).

\paragraph*{Emotional} Emotional engagement is more prevalent when engineers assume the role of authors. Reviewers are not always aware of the impact of their delivery styles. For instance, when we asked P12 and P18 to clarify their intention to maintain a positive tone in their comments, P12 responded, \emph{``I think I don't know whether it's something that I do consciously,''} and P18, \emph{``I try to be as much as possible neutral,''} respectively.

When processing feedback, engineers may experience a range of positive and negative emotions, contingent on the content and the delivery styles used by reviewers. For example, P2 expressed a feeling of ``inspiration'' and support in response to P19's constructive and invested feedback, \emph{``I was overwhelmed with P19 I liked the way that they formulated the code review. It was easy to read. I liked the way that they I think for me ... It inspires me sometimes with some of the colleagues, they go above and beyond''} (P2).

However, P6 had a different emotional experience with P11's feedback, as he perceived it as \emph{``brutal''}, causing a perceived feeling of an \emph{``attack.''} He explained: \emph{``... there were some feedback where I felt like, just being very brutal ... I do think that there was a tone that was a bit like off and brutal ... I just felt like it was kind of like attacking me''} (P6).

To alleviate some of the emotional taxation of negative feedback, engineers resort to emotional regulation or draw upon their personal values. Some of the emotional regulation strategies reported in our data include consciously avoiding negative reactions (e.g., P4), ``to not taking it personally`` (e.g., P10, P11), and mindfulness (e.g., P12). Software engineers also turn to their personal values to guide their behavior to navigate challenging feedback situations. Some of these values are ``avoiding hostility'' (e.g., P10), avoiding ``aggression'' (e.g., P16), and non-confrontational approach (e.g., P5, P7, P11). While emotional regulations are strategies used to manage individual emotional responses, personal values are guiding principles that shape long-term behavior and decision-making during challenging situations \cite{gross1998emerging,schwartz1992universals}.

Compared to LLM-assisted review, the engagement is less taxing emotionally than that of peers, as previously mentioned. P6 explains: \emph{``I feel like a tone goes a long way as well in terms of, like how you take feedback ... ChatGPT in that regards is better. I reacted positively to its tone''} (P6).

\paragraph*{Behavioral} While emotional engagement involves the internal feelings and affective responses elicited by feedback, behavioral engagement refers to the observable actions of individuals in response to feedback as a social stimulus \cite{fredricks2004school}. It encompasses how software engineers respond to feedback, including their willingness to seek clarification and proactively implement it. We assessed behavioral engagement in our data through expressed actions like initiating further dialogue with the reviewers or implementing changes to code \cite{skinner2009motivational,kahu2013framing}.

In reference to P2 and P6’s experiences above, their emotional responses to feedback influenced their subsequent behaviors. P2’s behavioral engagement seems generally positive, marked by reflection and increased effort. P6’s engagement appeared also positive, yet mixed, wanting more clarification and showing defensiveness. P2's actions show steps taken to revise and reflect on the code as a response to P19 feedback. He explained: \emph{``I went back to the code, and then I looked at the feedback, and then I looked back at the code thinking, okay, I see what they mean now''} (P2). This behavior may be more conducive to constructive improvement of the code, e.g., \emph{``you’re inspired to do well''} (P2). The absence of negative emotional impact allowed P2 to focus directly on the content rather than its delivery.

While P6 still engaged positively by seeking clarifications, his engagement may appear less efficient. Further dialogue mixed with defensive feelings, may delay and complicate the feedback loop and potential improvements to the code. P6 mentioned: \emph{``try to engage them... to understand what they mean by that review''} (P6), indicating delayed actions, compared to P2, whose emotional alignment seems to facilitate immediate actions to improve. Similar experience was echoed by P7, whose code was also reviewed by P11. He described her feedback as \emph{``little bit harsh''}, yet willing to engage in a constructive discussion to resolve the comments, \emph{``I would just have discussion with the reviewer ... we might come to a mutual agreement that would work for both of us ... I wouldn't just ignore''} (P7).

In comparison with an LLM-assisted review, the engineers in our sample show willingness to engage constructively with the LLM feedback, with a varying level of reservation, citing trust in LLMs, and lack of specific codebase context, as constraints. Despite the reservations, our interviewees seem to undertake a similar sense-making process of the feedback as for their peers. For example, while P18 has a clear preference for the LLM review, i.e., \emph{``ChatGPT had done the best ... if I have to choose one of the four, including chatGPT, yes, I would choose chatGPT''} (P18), he still suggests combining peers' with LLM reviews to mitigate the lack of context. He explained his reasoning: \emph{``AI might not be completely aware of what the context of the code you are writing ... what are your future goals, so how the code base should be''} (P18).

\paragraph*{Sense-making process} 

Sense-making is an active process of reflective evaluation of the feedback to pursue its adoption. Software engineers reflectively evaluate the feedback, then decide whether further dialogue with the reviewer is necessary before adoption. P5 summarizes this process: \emph{``I went through each one line by line, or suggestion by suggestion ... and sort of evaluated it, and if I agree with it or not, some of them I did, some of them I didn't. For the ones that I did, I would have no problem making those changes. There were some good suggestions in there. Some of the ones that I didn't necessarily agree with or I had questions about ... I would open up a dialog with the other developer and sort of have a back and forth discussion with them about the suggestion''} (P5).

The reflective evaluation reconciles the external feedback with the software engineer's internal knowledge structures. P8 and P9 describe it as \emph{``make sense to me''}, and \emph{``does it make any sense''}, respectively. Then, if this sense-making is successful and perceived to contribute to improvements, then the engineers apply the feedback, e.g., \emph{``If it does make any sense ... if I feel like they are more they're going to improve the readability and improve the style of coding. I'll go with that''} (P9).

Software engineers in our sample do not seem to discriminate in dealing with feedback, irrespective of the source, LLM or peers. They claim to undertake a similar sense-making process to evaluate and adopt LLM-generated feedback, despite its verbosity and high content, requiring more cognitive effort. For instance, P7 was asked about his decision-making process for ChatGPT-generate feedback, he replied, \emph{``decision making, I still follow the same process I use with my peers''} (P7).

\subsection{Reviewer context}

\noindent Reviewer context are the inherent attributes or characteristics of the reviewer. For humans, it includes seniority and familiarity with the reviewer. While seniority exerts authority to adopt the feedback, familiarity reduces the emotional impact. P13 explains how he would process feedback from his peers: \emph{``It depends who it's from. If it's from somebody who's more senior than me ... they know what they're talking about. If it was somebody not so senior, and maybe argue a bit''} (P13). This testimony shows that P13 attributes more credibility to his seniors' feedback compared to juniors.

Familiarity with peers fosters a sense of safety, reducing emotional tension because of the already existing relational capital. P15 explains: \emph{``if the review comes from someone I know in real life, this kind of interaction also helps to digest the feedback and improve your personal relationship, whether good or bad review ... I know that person cares''} (P15).

For LLMs, trust in LLM abilities and lacking the codebase context in the review constrain engineers willingness to adopt LLM's review. P20 said: \emph{``my experience, ChatGPT ... if you go a bit deeper in more advanced topics, it doesn't really know what to do clearly ... So I wouldn't trust ChatGPT''} (P20).

In this ``human vs. machine'' discovery, we learned that the introduction of LLMs brings a trade-off between cognitive and emotional engagement. Human reviewers have varied delivery styles, which may evoke stronger emotional responses, either positive or negative, influencing the process of acting upon the feedback. In contrast, LLM-assisted reviews, while more cognitively demanding, offer a consistent emotional experience. The resolution lies in engineers’ personal skills and values to cope and their sense-making process, to decide, and adopt the feedback.

\subsection{Human-AI collaboration}

There was a clear divergence in preferences regarding how feedback should be delivered amongst the engineers in our sample. Some engineers preferred concise, actionable suggestions (e.g., P17, P18, and P20), while others wanted more detailed and educational feedback (e.g., P11 and P12). For example, P5 described his preference for LLM-generated feedback to have a high signal-to-noise ratio, containing a high proportion of useful, relevant, and actionable information (the ``signal'') relative to irrelevant or distracting content (the ``noise''). He said: \emph{``... there is a lot of information ... any point in here could be valid and worth considering, but I'm just like the signal to noise ratio, how much in here is like, truly valuable information that I would put a lot of thought into considering that's my concern''} (P5).

Willingness to adopt LLM in code review is marked by tension between the efficiency and technical completeness of the LLM-generated feedback and the desire for the human ``touch'' (P5) and expertise. When P12 was asked if she prefers P2 or ChatGPT reviews, she conveyed a strong preference: \emph{``Well, ChatGPT being a faceless chat box is probably a factor. But what I've gathered from working with other people, especially people who are senior developers... they can give me more insights and things that I haven't previously seen and still connect them to the things that I'm currently working on. That I think is super valuable insight for me''} (P12). P15 echoes this tension in this statement: \emph{``I can say ChatGPT one because it's more complete. So it found all the errors and all the things that I can improve. On the other hand, I can say I found the other participant's reviews more, let's say heartwarming, because some other person spent this time to read my code, understand it, and then do a review ... [ChatGPT] spent so one second''} (P15).

To balance LLM's efficiency with the human elements and expertise, most of our participants suggested combining both peers' and LLM's reviews. For example, even though P13 is willing to replace P16 and P17 with ChatGPT, he seems not willing yet to let go of the humans in the loop. \emph{``I would replace two of them [i.e., P16 and P17] with ChatGPT, but I feel like if given the choice, I would use ChatGPT for a pre-review check. You could somehow get that to look over it while it's still in the IDE, make your changes, and then put it out for review and then let a person look at it''} (P13).

\section{Discussion}\label{sec:discussion}

\noindent In this section, we discuss the implications of our findings. Three main takeaways from this study: \emph{Emotional intelligence (EI)}, \emph{feedback constructiveness}, and \emph{human-AI collaboration in SE}. The human review remains highly relevant, but the emotional impact is significant. EI has the potential to alleviate this emotional toll for an efficient process and the well-being of engineers. Our findings show diverse preferences for AI interaction, mainly the formulation of the feedback, as well as a willingness among engineers to adopt them. To capitalize on their capabilities, future AI tools should adapt to individual preferences.

\subsection{Emotional intelligence}

\noindent Our findings indicate that software engineers use different coping mechanisms to mitigate the emotional cost of engaging with peers' feedback. Some of these mechanisms are builtin and stem from their personal values (e.g., non-confrontational approach) or strategies (e.g., mindfulness) that they have developed. This implies the importance of fostering emotional intelligence amongst software engineers through training and support systems.

Humans are humans, and developers are humans, too! The expectation that an individual will perform consistently and constructively is perhaps unrealistic. Given the emotional variability inherent in human behaviors \cite{lazarus1984stress}, fostering emotional intelligence is essential to help engineers navigate the psychological challenges posed when engaging with peer feedback \cite{lazarus1984stress,connor2003development}. This built-in emotional intelligence has the potential to minimize frictions in the review process and shift the focus to code improvements, a primary objective of code review.

Significant work has been devoted to emotions in SE \cite{novielli2019sentiment,sanchez2019taking}. Amongst others, topics range from emotions in requirements engineering, e.g., \cite{colomo2010study,madampe2023framework}, emotions in software artifacts, e.g., \cite{murgia2014developers}, developer's emotions, e.g., \cite{murgia2014developers,girardi2021emotions}, and the impact of emotions on software quality artifacts \cite{khan2021understanding}. However, S\'anchez-Gord\'on and Colomo-Palacios systematic review show that managing emotions is less explored \cite{sanchez2019taking}, despite its potential to inform decision-making and social interactions \cite{goleman2020emotional}.

Emotional intelligence (EI) is the individual ability to recognize, understand, and proactive manage emotions to navigate emotional situations in social interactions and subsequent decision-making \cite{goleman2020emotional}.

Kosti et al. reported that emotional intelligence influences work preferences among software engineers \cite{kosti2014personality}. They suggest that software engineers with higher emotional intelligence prefer being responsible for the entire development process and autonomy in prioritizing their tasks \cite{kosti2014personality}. Rezvani and Khosravi found that EI mitigates stress and fosters trust among software developers, which results in increased performance \cite{rezvani2019emotional}.

Madampe et al. reported twelve EI strategies used by SE practitioners during requirements change handling \cite{madampe2023framework}. These strategies have a direct impact on self-regulation of emotions and relationships. They range from peer support, openness in communication, and ``social rituals'' \cite{madampe2023framework}. As our findings suggest, coping mechanisms are rooted in personal values like non-confrontational approaches or strategies like mindfulness, which help engineers to alleviate the emotional costs of feedback. These mechanisms can be viewed as manifestations of EI regulation \cite{bar2006bar}.

Hodzic et al. suggest that EI ``should be considered effective interventions'' \cite{hodzic2018efficient,mattingly2019can}. This can be done through support programs within the organization or direct training \cite{hodzic2018efficient,mattingly2019can}. Therefore, we propose this implication:

\textbf{\textit{Implication \#1:}} Given its potential to minimize friction and contribute to more effective code reviews in the pursuit of improved code quality, EI training should become part of standard organizational support and development programs in SE.

\subsection{Feedback constructiveness}

Our findings suggest that constructive feedback requires lower cognitive demand and positive emotional responses. Gon\c{c}alves et al. report that negative feedback in code review causes interpersonal conflicts \cite{wurzel2024constructive}. They suggest that developers should constructively manage their conflicts to foster a more positive code review environment \cite{wurzel2024constructive}. Our findings show that the feedback delivery itself triggers negative emotional responses prior to further dialogue being initiated to resolve the comments. Hence, we suggest promoting feedback constructiveness to mitigate negative emotional responses and subsequent conflicts.

Kluger and DeNisi note that feedback content and delivery have varying effects on performance \cite{kluger1996effects}. Constructive feedback is the most effective, as it focuses on the task and improvements, which minimizes negative emotional responses \cite{kluger1996effects}. Hattie and Timperley suggest that feedback is constructive when it meets the criteria of being clear, task-oriented, and specific \cite{hattie2007power}. In educational settings, Carless suggests that feedback should aim to correct errors, foster a growth mindset, and encourage positive emotions to support learning \cite{carless2006differing}. Feedback should also be framed in a way that promotes cognitive engagement and reduces emotional defensiveness \cite{shute2008focus}.

\textbf{Implication \#2:} Feedback constructiveness should be part of computer science and software engineering educational programs. Organizations and SE teams should also document, promote, and reward constructive feedback.

\subsection{Human-AI collaboration in SE}

The transition from human-human to human-AI collaboration brings new complexities. Our findings show that while software engineers are generally willing to adopt LLM-assisted code review, there are varying preferences for how AI delivers feedback and an apparent tension between the efficiency and technical completeness of AI and the desire for the human ``touch'' and expertise.

The Sowa et al. study on human-AI collaboration in knowledge work suggests that when human-AI collaboration is enhanced, productivity increases \cite{sowa2021cobots}. They conclude that AI automation should rather shift focus to collaborative approaches where humans and AI work closely together \cite{sowa2021cobots}, balancing AI’s computational strengths with the relational and experiential aspects of humans.

Studies in human-AI collaboration specific to traditional SE practices and processes remain scarce \cite{hamza2024human}. Hamza et al. study suggests that software engineers expect a ``collaborative partner'' not merely a tool. They further explain that this partnership is also expected to be effective and dialogic to ensure accurate and relevant AI output \cite{hamza2024human}.

To ensure effective engagement, our findings indicate that willingness to adopt AI is characterized by varying preferences for interaction. In the context of LLM-assisted code review, while some software engineers prefer concise, actionable feedback that enables faster decision-making and implementation, others appreciate more detailed, educational feedback that contributes to their learning. We draw parallels with the Ghorbani et al. study on open-source contributors' preferences for GitHub bot interactions \cite{ghorbani2023autonomy}. They found that developers prefer adaptable and customizable bot interactions, catering to individual preferences \cite{ghorbani2023autonomy}.

Jang et al. report that autistic worker preferred LLMs for their structured, clear responses and the privacy they offered, which is in contrast with the emotional risks associated with seeking help from humans \cite{jang2024s}. AI can play a moderator role to make human feedback more constructive with a positive tone.
 
\textbf{Implication \#3:} To capitalize on the potential of AI tools in SE processes, future AI tools should incorporate personalization features that align feedback delivery with individual preferences. In addition, AI can play the role of a moderator, adjusting or enhancing the tone and constructiveness of feedback to reduce the emotional cost and foster developer well-being.

Although our implications do not lead to concrete and actionable guidelines, they offer valuable directions for future research. For example, our implication for EI training does not specify exactly how training for EI is SE may be implemented. Future work could draw from the extensive work on psychology to develop tailored interventions and strategies for fostering emotional intelligence in software engineering contexts. Similarly, our implication on feedback constructiveness raises important questions about how to promote more effective and emotionally attuned feedback in a human-led review. Our findings on human-AI collaboration also open avenues for further research on what the future AI-assisted SE may look like and how to define this ``AI partnership.''

\section{Trustworthiness}\label{sec:validity}

\noindent To ensure our study's trustworthiness \cite{miles2014qualitative}, we adopted these methods: \textit{saturation}, \textit{peer debriefing}, we provide a \textit{thick description} (see Sect. \ref{sec:methods} for these methods), and we conducted a \textit{member checking} activity.

\textit{Member checking}: We also carried out member checking \cite{birt2016member} to seek feedback from our participants. This method allowed us to receive feedback on whether the analysis accurately represented our participants' experiences and perspectives, thereby adding a layer of verification to the accuracy of our interpretations \cite{birt2016member}. We documented our preliminary findings in survey-like page forms and invited all participants to take part and comment on our interpretations, and 19 from 20 responded. Most comments were supportive of our interpretations and, in some instances, augmented our findings with additional data in the comments the participant provided. We paid an additional \textsterling5 for this effort. See Sect.\ref{sec:replication} for supporting material and data.

\section{Limitations and Trade-offs}\label{sec:limit}

\noindent We conducted the code review process anonymously. We were constrained by the Prolific requirement to maintain participant anonymity throughout the study, whic is not the case in open source and proprietary software development. In the latter, developers are colleagues and interact directly on a daily basis. In such a social context, developers may align the tone of their feedback according to status, relationships, and team culture \cite{bird2006mining,kononenko2016code}. Similarly, in open-source development, communities adopt different styles and strategies in pull request reviews, from ``lenient'' to ``protective'' according to their history, culture, and sustainability strategy \cite{alami2020foss,alami2021pull}. Aware of this limitation, we asked during the interview whether this factor has influenced our participants approach to reviews. Most of our participants claimed that was not the case.

The code reviews were conducted in an artificial setting characteristic of a research environment. Such a controlled environment may not capture the inherent complexity of a real-world setting. However, this controlled environment allowed us to focus on the variables of interest and gain nuanced findings \cite{stol2018abc}. We also acknowledge that our participants may have modified their approach to the review compared to a professional setting, given the research nature of their contribution. However, in the interview, we prompted all our participants to explain the content and the delivery style they used. This ``reflective questioning'' uncovered the ``authentic'' reasoning even when the original actions occurred in artificial settings \cite{silverman2013counts}.

\section{Conclusion}\label{sec:conclusion}

\noindent In this study, we learned that in peer-led review, engagement is multi-dimensional, covering cognitive, emotional, and behavioral responses. Software engineers engage in a sense-making process before making a decision about the feedback suggestions. In an LLM-assisted review, the emotional toll is alleviated by the consistent positive delivery; however, the cognitive load to process the feedback is higher. LLM were perceived positively by software engineers for their technical superiority and completeness. However, their adoption seem to be contingent to diverse preferences on the feedback delivery.

\paragraph*{Acknowledgment}

We thank all the study participants for their time and effort. This study was funded by the department of computer science at Aalborg University; research funding for tenure-track assistant professors.

\bibliographystyle{IEEETrans}
\bibliography{references}

\end{document}